\documentstyle{mn}
\newread\epsffilein    
\newif\ifepsffileok    
\newif\ifepsfbbfound   
\newif\ifepsfverbose   
\newdimen\epsfxsize    
\newdimen\epsfysize    
\newdimen\epsftsize    
\newdimen\epsfrsize    
\newdimen\epsftmp      
\newdimen\pspoints     
\def\epsfbox#1{\global\def\epsfllx{72}\global\def\epsflly{72}%
   \global\def\epsfurx{540}\global\def\epsfury{720}%
   \def\lbracket{[}\def\testit{#1}\ifx\testit\lbracket
   \let\next=\epsfgetlitbb\else\let\next=\epsfnormal\fi\next{#1}}%
\def\epsfgetlitbb#1#2 #3 #4 #5]#6{\epsfgrab #2 #3 #4 #5 .\\%
   \epsfsetgraph{#6}}%
\def\epsfnormal#1{\epsfgetbb{#1}\epsfsetgraph{#1}}%
\def\epsfgetbb#1{%
\openin\epsffilein=#1
\ifeof\epsffilein\errmessage{I couldn't open #1, will ignore it}\else
   {\epsffileoktrue \chardef\other=12
    \def\do##1{\catcode`##1=\other}\dospecials \catcode`\ =10
    \loop
       \read\epsffilein to \epsffileline
       \ifeof\epsffilein\epsffileokfalse\else
          \expandafter\epsfaux\epsffileline:. \\%
       \fi
   \ifepsffileok\repeat
   \ifepsfbbfound\else
    \ifepsfverbose\message{No bounding box comment in #1; using defaults}\fi\fi
   }\closein\epsffilein\fi}%
\def\epsfsetgraph#1{%
   \epsfrsize=\epsfury\pspoints
   \advance\epsfrsize by-\epsflly\pspoints
   \epsftsize=\epsfurx\pspoints
   \advance\epsftsize by-\epsfllx\pspoints
   \epsfxsize\epsfsize\epsftsize\epsfrsize
   \ifnum\epsfxsize=0 \ifnum\epsfysize=0
      \epsfxsize=\epsftsize \epsfysize=\epsfrsize
     \else\epsftmp=\epsftsize \divide\epsftmp\epsfrsize
       \epsfxsize=\epsfysize \multiply\epsfxsize\epsftmp
       \multiply\epsftmp\epsfrsize \advance\epsftsize-\epsftmp
       \epsftmp=\epsfysize
       \loop \advance\epsftsize\epsftsize \divide\epsftmp 2
       \ifnum\epsftmp>0
          \ifnum\epsftsize<\epsfrsize\else
             \advance\epsftsize-\epsfrsize \advance\epsfxsize\epsftmp \fi
       \repeat
     \fi
   \else\epsftmp=\epsfrsize \divide\epsftmp\epsftsize
     \epsfysize=\epsfxsize \multiply\epsfysize\epsftmp   
     \multiply\epsftmp\epsftsize \advance\epsfrsize-\epsftmp
     \epsftmp=\epsfxsize
     \loop \advance\epsfrsize\epsfrsize \divide\epsftmp 2
     \ifnum\epsftmp>0
        \ifnum\epsfrsize<\epsftsize\else
           \advance\epsfrsize-\epsftsize \advance\epsfysize\epsftmp \fi
     \repeat     
   \fi
   \ifepsfverbose\message{#1: width=\the\epsfxsize, height=\the\epsfysize}\fi
   \epsftmp=10\epsfxsize \divide\epsftmp\pspoints
   \newcount\figskipcount
      \message{#1 \the\epsfysize  }
   \vbox to\epsfysize{\vfil\hbox to\epsfxsize{%
      \includegraphics{#1}%
      \hfil}}%
\epsfxsize=0pt\epsfysize=0pt}%

{\catcode`\%=12 \global\let\epsfpercent=
\long\def\epsfaux#1#2:#3\\{\ifx#1\epsfpercent
   \def\testit{#2}\ifx\testit\epsfbblit
      \epsfgrab #3 . . . \\%
      \epsffileokfalse
      \global\epsfbbfoundtrue
   \fi\else\ifx#1\par\else\epsffileokfalse\fi\fi}%
\def\epsfgrab #1 #2 #3 #4 #5\\{%
   \global\def\epsfllx{#1}\ifx\epsfllx\empty
      \epsfgrab #2 #3 #4 #5 .\\\else
   \global\def\epsflly{#2}%
   \global\def\epsfurx{#3}\global\def\epsfury{#4}\fi}%
\def\epsfsize#1#2{\epsfxsize}
\def\figinsert#1#2{\epsfbox{#1} \message{#2} }    
\title[The Durham/UKST Galaxy Redshift Survey]
      {The Durham/UKST Galaxy Redshift Survey - II.\\ The Field Galaxy
Luminosity Function.}
\author[A. Ratcliffe et al.]
       {A. Ratcliffe$^{1}$, T. Shanks$^{1}$, Q.A. Parker$^{2}$ and R.
Fong$^{1}$ \\
        $^{1}$Physics Deptartment, University of Durham, South Road, Durham,
DH1 3LE.\\
	$^{2}$Anglo-Australian Observatory, Coonabarabran, NSW 2357,
Australia.}
\begin{document}

\maketitle

\begin{abstract}
We present the results for the galaxy luminosity function as estimated
from the Durham/UKST Galaxy Redshift Survey. This survey is magnitude
limited to \mbox{$b_{J} \sim 17$}, contains $\sim$2500 galaxies sampled
at a rate of one in three and surveys a $\sim$$4 \times 10^{6}
(h^{-1}$Mpc$)^{3}$ volume of space. The maximum likelihood parameters
for a standard Schechter luminosity function are estimated to be
$M^{*}_{b_{J}} = -19.72 \pm 0.09$, $\alpha = -1.14 \pm 0.08$ and
$\phi^{*} = (1.2 \pm 0.2) \times 10^{-2}$ ($h^{3}$Mpc$^{-3}$).
Attempting to correct for the scatter in the observed magnitudes leads
to a flatter faint end slope, $\alpha = -1.04 \pm 0.08$, which,
combined with the different luminosity function shape, causes a higher
normalisation to be estimated, $\phi^{*} = (1.7 \pm 0.3) \times
10^{-2}$ ($h^{3}$Mpc$^{-3}$). Neither of these parametric functions
provides a good formal fit to the non-parametric estimate of the
luminosity function. A comparison with galaxy luminosity functions from
other redshift surveys shows good agreement and the shape of the
luminosity function now appears well-defined down to $M_{b_{J}} \simeq
-17$. There are some discrepancies between the different surveys for
galaxies fainter than this absolute magnitude. However, our estimate
agrees well with that from the APM-Stromlo Galaxy Redshift Survey and
we measure a fairly flat faint end slope.
\end{abstract}


\section{Introduction}

The galaxy luminosity function is a fundamental quantity in
observational cosmology. Indeed, a knowledge of its form it is
essential for an accurate analysis of galaxy clustering in magnitude
limited redshift surveys (e.g. Efstathiou 1988) and also for a proper
interpretation of the observed galaxy number-magnitude counts (e.g.
Metcalfe et al. 1995a). On the theoretical side, the luminosity
function is one of the first tests that models of galaxy formation and
evolution must pass in order to be called successful (e.g. Cole et al.
1994).

Current interest in the observed galaxy luminosity function takes the
form of an accurate determination of the faint end slope and overall
normalisation (e.g. Marzke et al. 1994; Vettolani et al. 1996), the
relationship between the field and cluster luminosity functions (e.g.
Driver et al. 1994; Kashikawa et al. 1995) and the evolution with
redshift of the luminosity function (Lilly et al. 1995; Ellis et al.
1996). Meanwhile, theoreticians are attempting to improve the
sophistication of their modelling in order to predict more realistic
luminosity functions (e.g. Frenk et al. 1996).

The initial clustering results, redshift maps, etc. of the Durham/UKST
Galaxy Redshift Survey were summarized in the first paper of this
series \cite{mymnras}. In this paper we present a detailed analysis of
the luminosity function and space density of galaxies in this optically
selected survey. We briefly describe our survey in
Section~\ref{redsursec}. The techniques used to estimate the luminosity
function, radial density and normalisation are outlined in
Section~\ref{methsec}. The results from the Durham/UKST survey are then
presented in Section~\ref{ressec}. Section~\ref{disssec} compares and
discusses the current state of the galaxy luminosity function. Finally,
in Section~\ref{concsec} we summarize our conclusions from this
analysis.

\section{The Durham/UKST Galaxy Redshift Survey} \label{redsursec}

The Durham/UKST Galaxy Redshift Survey was constructed using the FLAIR
fibre optic system \cite{FLAIR} on the 1.2m UK Schmidt Telescope at
Siding Spring, Australia. This survey is based on the astrometry and
photometry from the Edinburgh/Durham Southern Galaxy Catalogue (EDSGC;
Collins, Heydon-Dumbleton \& MacGillivray 1988; Collins, Nichol \&
Lumsden 1992) and was completed in 1995 after a 3-yr observing
programme. The survey itself covers a $\sim$$20^{\circ} \times
75^{\circ}$ area centered on the South Galactic Pole (60 UKST plates)
and is sparse sampled at a rate of one in three of the galaxies to
$b_{J} \simeq 17$ mag. The resulting survey contains $\sim$2500
redshifts, probes to a depth greater than $300h^{-1}$Mpc, with a median
depth of $\sim$$150h^{-1}$Mpc, and surveys a volume of space $\sim$$4
\times 10^{6} (h^{-1}$Mpc$)^{3}$.

The survey is $>$75 per cent complete to the nominal magnitude limit of
$b_{J} = 17.0$ mag. This incompleteness was mainly caused by poor
observing conditions, intrinsically low throughput fibres and other
various observational effects. In a comparison with $\sim$150 published
galaxy velocities (Peterson et al. 1986; Fairall \& Jones 1988;
Metcalfe et al. 1989; da Costa et al. 1991) our measured redshifts had
negligible offset and were accurate to $\pm 150$ kms$^{-1}$. The
scatter in the EDSGC magnitudes has been estimated at $\pm 0.22$ mags
\cite{nmb} for a sample of $\sim$100 galaxies. This scatter has been
confirmed by a preliminary analysis of a larger sample of high quality
CCD photometry. All of these observational details are discussed
further in a forthcoming data paper (Ratcliffe et al., in
preparation).

\section{Estimating the Luminosity Function, Radial Density and
Normalisation} \label{methsec}

We have estimated the galaxy luminosity function, $\phi(L)$, from the
Durham/UKST Galaxy Redshift Survey firstly using the very basic and
common method of Schmidt (1968) and secondly using two maximum
likelihood techniques; the parametric method proposed by Sandage,
Tammann \& Yahil (1979) and the non-parametric stepwise maximum
likelihood method of Efstathiou, Ellis \& Peterson (1988). These
methods have well defined error properties and the maximum likelihood
techniques have been constructed such that they are unbiased by density
inhomogeneities in the galaxy distribution. They assume that the
luminosity function has a universal form and hence the number density
is separable into a product of functions of luminosity and position,
$n(L,{\bf r}) = \phi(L)\rho({\bf r})$. Unfortunately, in using this
assumption all the density and normalisation information is lost. To
obtain the radial density information one uses a similar technique to
the above non-parametric method as proposed by Saunders et al. (1990).
To determine the absolute normalisation the minimum variance iterative
technique of Loveday et al. (1992) is employed. This is a development
of the method originally proposed by Davis \& Huchra (1982).

\subsection{The Luminosity Function} \label{lumfnmethsec}

Probably the most basic estimator of the luminosity function is the
`$1/V_{max}$' method due to Schmidt (1968)
\begin{equation}
\phi(L)dL = \sum_{i} \frac{1}{V_{max}(L_{i})} ,
\end{equation}
where the sum extends over all galaxies in the $L \rightarrow L + dL$
luminosity interval and $V_{max}(L_{i})$ is the maximum volume that the
galaxy of luminosity $L_{i}$ could be seen in (given the survey's
physical and apparent magnitude limits). Unfortunately, this estimate
is biased by density inhomogeneities in the galaxy distribution but is
still useful for initial comparisons. Error properties of this method
can be estimated by a simple rms of the galaxies in each luminosity
interval in question
\begin{equation}
{\rm Var}(\phi) = \sum_{i} \frac{1}{V^{2}_{max}(L_{i})} ,
\label{vmaxerr}
\end{equation}
For future reference this will be called the VMAX method.

Methods that are independent of the galaxy density inhomogeneities can
be constructed as follows. One can form a likelihood, $\cal L$, based
on the probability of observing a galaxy of luminosity $L_{i}$ at
redshift $z_{i}$ in a magnitude limited redshift survey
\begin{equation}
p_{i} \propto \phi(L_{i}) \left/ \int_{L_{min}(z_{i})}^{\infty}
\phi(L)dL \right. , \label{prob}
\end{equation}
where
\begin{equation}
{\cal L} = \prod_{i=1}^{N} p_{i} . \label{like}
\end{equation}
The product extends over all of the $N$ galaxies in the survey and
$L_{min}(z_{i})$ is the minimum absolute luminosity that a galaxy at
redshift $z_{i}$ could have and still be included in the survey. A
maximum absolute luminosity could also be incorporated into
equation~\ref{prob}, altering the upper limit of the integral. In
practice this makes little difference to any results. The best estimate
of $\phi(L)$ is then given when $\cal L$ (or equivalently $\ln {\cal
L}$) is maximised.

In the parametric case one assumes a function form for $\phi(L)$ and
maximises equation~\ref{like} with respect to the parameters of this
assumed functional form \cite{sty}. In keeping with tradition we use
the `Schechter function' \cite{schechter} to describe $\phi(L)$
\begin{equation}
\phi(L)dL = \phi^{*} \left(\frac{L}{L^{*}}\right)^{\alpha} \exp
\left(-\frac{L}{L^{*}}\right) d\left(\frac{L}{L^{*}}\right) ,
\label{sch}
\end{equation}
where the three parameters are a normalisation, $\phi^{*}$, a faint end
slope, $\alpha$, and a characteristic luminosity, $L^{*}$, (or
equivalently absolute magnitude, $M^{*}$). In practice $\phi^{*}$
cancels from equation~\ref{prob} and a maximum is determined in the
($M^{*}, \alpha$) likelihood space. A further consideration is the
effect of the observed scatter in the measured magnitudes. Efstathiou
et al. (1988) modelled this by approximating these errors with a
Gaussian distribution of zero mean and $\sigma$ rms. The observed
(convolved) Schechter function, $\phi_{o}$, then becomes
\begin{equation}
\phi_{o}(M) = \frac{1}{\sqrt{2\pi}\sigma} \int_{-\infty}^{\infty}
\phi(M^{\prime}) \exp{\left[-\frac{1}{2\sigma^{2}}(M^{\prime} -
M)^{2}\right]} dM^{\prime} , \label{consch}
\end{equation}
Given that equation~\ref{consch} is a slightly naive approximation it
is not obvious that it will substantially improve the quality of the
fit. Finally, error properties of this method can be estimated by the
deviations of ${\cal L}$ from its maximum value. One can plot error
ellipses in the ($M^{*}, \alpha$) plane by considering the contours
corresponding to
\begin{equation}
\ln {\cal L} = \ln {\cal L}_{max} - \frac{1}{2}\chi_{\beta}^{2}(n) ,
\label{styerr}
\end{equation}
where ${\cal L}_{max}$ is the maximum likelihood, $n$ is the number of
free parameters (namely two, $\alpha$ and $M^{*}$) and $\beta$ is the
required confidence level for that number of free parameters. For
future reference this will be called the STY method.

In the non-parametric case one assumes that $\phi(M)$ can be written as
a series of constant steps across given luminosity intervals and one
maximises to find the relative values of these steps \cite{swml}
\begin{eqnarray}
\phi(M) = \phi_{k}, & \left| M - M_{k} \right| \leq \Delta M/2, & k =
1,\ldots ,N_{p} ,
\end{eqnarray}
where $N_{p}$ is the number of constant steps (or bins) of width
$\Delta M$. The maximisation condition $(\frac{\partial \ln {\cal
L}}{\partial \phi_{k}} = 0)$ quickly leads to an equation that can be
solved iteratively
\begin{equation}
\phi_{k} = \frac{\sum_{i=1}^{N}W(M_{i}-M_{k})}{\sum_{i=1}^{N} \left[
\frac{H(M_{k}-M_{min}(z_{i}))\Delta M}{\sum_{j=1}^{N_{p}}
H(M_{j}-M_{min}(z_{i}))\phi_{j}\Delta M} \right]} , \label{swml}
\end{equation}
where
\begin{equation}
W(x) = \left\{ \begin{array}{ll}
1 & \left| x \right| \leq \Delta M/2 \\
0 & {\rm otherwise}
\end{array} \right. , \label{sumw}
\end{equation}
and 
\begin{equation}
H(x) = \left\{ \begin{array}{lcl}
0 & & \,\, x \;\; \leq \;\; -\Delta M/2 \\
\frac{1}{2} + \frac{x}{\Delta M} & & \left| x \right| \,\, \leq
\;\;\;\;\; \Delta M/2 \\
1 & & \,\, x \;\; \geq \;\;\;\;\; \Delta M/2
\end{array} \right. .
\end{equation}
Finally, error properties of the $\phi_{k}$'s can be estimated from the
covariance matrix
\begin{equation}
{\rm Cov}(\phi_{k}) = - \left(\frac{\partial^{2}\ln {\cal L}}{\partial
\phi_{l}^{2}}\right)_{\phi_{l}=\phi_{k}}^{-1} . \label{covmat}
\end{equation}
A constraint involving the $\phi_{k}$'s is usually introduced into the
likelihood equation in order to ensure that the information matrix is
non-singular and hence invertible as is assumed in equation~\ref{covmat}.
This constraint does not affect the shape of the maximum likelihood
solution. We take a simple approach and fix one of the maximum
likelihood $\phi_{k}$'s to be constant. We then assume that one can
neglect the off-diagonal elements \cite{iras}. Using this assumption
gives
\begin{eqnarray*}
{\rm Var}(\phi_{k}) & = & \left( \sum_{i=1}^{N}
\left[\frac{W(M_{i}-M_{k})}{\phi_{k}^{2}} \right] \right. \\
& - & \left. \sum_{i=1}^{N}
\left[\frac{ H(M_{k}-M_{min}(z_{i}))\Delta M}{\sum_{j=1}^{N_{p}}
H(M_{j}-M_{min}(z_{i}))\phi_{j}\Delta M} \right]^{2}  \right)^{-1} .
\label{swmlerr}
\end{eqnarray*}
In practice the cross-derivatives are small and therefore our
approximation will only slightly underestimate the errors in the
$\phi_{k}$. For future reference this will be called the SWML method.

A major problem with the STY method is that it will always return a
maximum likelihood solution regardless of the assumed parametric form
and how well it represents the actual luminosity function. Therefore,
it is necessary to test the goodness of fit of this solution with a
likelihood ratio test. We assume that the non-parametric SWML method
provides a good representation of the actual luminosity function shape.
Let ${\cal L}_{1}$ be the likelihood calculated using the maximum
likelihood solution of the given parametric form (STY) and let ${\cal
L}_{2}$ be the likelihood calculated using the maximum likelihood
solution of the $\phi_{k}$'s (SWML). Efstathiou et al. (1988) have
shown that $2\ln ({\cal L}_{1}/{\cal L}_{2})$ approximately behaves
like a $\chi^{2}$ statistic with $(N_{p}-3)$ degrees of freedom.
However, for the answer to be independent of the bin size and the
number of bins, the likelihood ${\cal L}_{1}$ should be calculated
using the SWML likelihood formula (which produced ${\cal L}_{2}$) with
the set of $\phi_{k}$'s from
\begin{equation}
\phi_{k} \simeq \frac{\int \phi(L)dN(L)} {\int dN(L)} \simeq \frac{\int
\phi(L)L^{\frac{3}{2}}dL} {\int L^{\frac{3}{2}}dL} . \label{likeratio}
\end{equation}
The integrals in equation~\ref{likeratio} are over the luminosity
interval in question, $[L_{k}-\frac{\Delta L}{2},L_{k}+\frac{\Delta
L}{2}]$ (Efstathiou et al. 1988; Saunders et al. 1990).

\subsection{The Radial Density} \label{raddenmethsec}

Using a similar method to Section~\ref{lumfnmethsec} one can estimate
the radial density function, $\rho(r)$. First consider the probability
of observing a galaxy of luminosity $L_{i}$ at redshift $z_{i}$ in a
magnitude limited redshift survey. As before a likelihood is formed
from the product of these probabilities
\begin{equation}
{\cal L} = \prod_{i=1}^{N}
\frac{\rho(z_{i})}{\int_{\max[z_{low},z_{min}(L_{i})]}^{\min[z_{hi},z_{max}(L_{i})]}
\rho(z_{i}) \left(\frac{dV}{dz}\right) dz} , \label{raddeneqn}
\end{equation}
where $z_{min}(L_{i})$ and $z_{max}(L_{i})$ are the minimum and maximum
redshifts at which a galaxy of luminosity $L_{i}$ could be seen and
still be included in the survey. Saunders et al. (1990) assume that one
can write $\rho(r)$ as a series of step functions in radial distance
and then solve for the maximum likelihood $\rho_{k}$'s by iteration as
before. Finally, error properties of the $\rho_{k}$'s can be estimated
from the appropriate covariance matrix and we use similar
approximations to those made in Section~\ref{lumfnmethsec} to simplify
the analysis involved.

\subsection{The Normalisation} \label{normmethsec}

By their method of construction these maximum likelihood techniques
cannot provide an overall normalisation. Therefore, one must adopt
other techniques to estimate the mean space density of galaxies,
$\bar{n}$, and the normalisation of the luminosity function,
$\phi^{*}$.

The expected distribution of the number of galaxies as a function of
distance~$r$ is given by
\begin{equation}
n(r) = f\bar{n} \; V(r) \; S(r) , \label{nbar1}
\end{equation}
where $f$ is the sampling rate of the survey, $\bar{n}$ is the mean
spatial galaxy density of the survey and $V(r)$ is the volume element
of the survey at a distance $r$. The radial selection function of the
survey, $S(r)$, is given by
\begin{equation}
S(r) =
\frac{\int_{\max[L_{low},L_{min}(r)]}^{\infty}\phi(L)dL}{\int_{L_{low}}^{\infty}\phi(L)dL}
, \label{selfn}
\end{equation}
where $L_{low}$ is some minimum absolute galaxy luminosity and
$L_{min}(r)$ was defined in Section~\ref{lumfnmethsec}. If a Schechter
function is assumed then the above integrals become incomplete Gamma
functions, $\Gamma\left(\alpha+1,x\right)$. The mean spatial galaxy
density is also related to the luminosity function by
\begin{equation}
\bar{n} = \int_{L_{low}}^{\infty}\phi(L)dL , \label{nbar2}
\end{equation}
and so using the luminosity function from either equation~\ref{sch}
or~\ref{consch} one finds a relationship involving $\bar{n}$,
$\phi^{*}$ and an integral over the shape of the assumed functional
form. Using this formalism Davis \& Huchra (1982) showed that an
unbiased estimator for $\bar{n}$ is given by
\begin{equation}
\bar{n} = \frac{\sum_{i=1}^{N} w(r_{i})/f}{\int_{r_{min}}^{r_{max}}
S(r)w(r)dV} , \label{nbar3}
\end{equation}
where $w(r)$ is any weight function we choose and $[r_{min} , r_{max}]$
is the distance range over which we use galaxies. One choice is to
weight all galaxies equally, whereby $w = 1$. However, Davis \& Huchra
(1982) showed that the minimum variance in $\bar{n}$ occurs when
\begin{equation}
w(r) = \frac{1}{1 + 4\pi f\bar{n} J_{3}(r_{c}) S(r)} , \label{weight}
\end{equation}
where
\begin{equation}
J_{3}(r_{c}) = \int_{0}^{r_{c}} x^{2}\xi(x)dx ,
\end{equation}
$r_{c}$ is the scale on which $J_{3}$ converges to a maximum value and
$\xi(x)$ is the galaxy 2-point correlation function. Given that this
weighting depends on the quantity we are interested in ($\bar{n}$),
Loveday et al. (1992) have developed an iterative scheme involving
equations~\ref{nbar3} and~\ref{weight}. This method should produce the
minimum variance estimate of $\bar{n}$ if $J_{3}(r_{c})$ converges on a
scale $r_{c}$ smaller than the survey. The variance of $\bar{n}$ is
given by \cite{dh}
\begin{equation}
{\rm Var}(\bar{n}) = \frac{\bar{n}\int w^{2}SdV + f\bar{n}^{2}\int
w_{1}w_{2}S_{1}S_{2}\xi(x_{12})dV_{1}dV_{2}}{f\left(\int
wSdV\right)^{2}} . \label{nbarerr}
\end{equation}
In practice the final answer for $\bar{n}$ (and hence $\phi^{*}$)
depends little on the value of $J_{3}(r_{c})$ used. Therefore, we adopt
the value $4 \pi J_{3}(r_{c}) = 5000 h^{-3}$Mpc$^{3}$ and doubling or
halving this only makes a few per cent difference to $\bar{n}$. Also,
we are free to choose the value of $L_{low}$ (or equivalently
$M_{low}$) in equations~\ref{selfn} and~\ref{nbar2} as long as we are
consistent and throw away all the fainter galaxies from the sum in the
numerator of equation~\ref{nbar3}. Obviously $\bar{n}$ is very
sensitive to the value of $M_{low}$ used but $\phi^{*}$ is very stable
over a range of $\sim$5 magnitudes.

\section{Results from the Durham/UKST Galaxy Redshift Survey}
\label{ressec}

We use galaxies in the distance range [5,350]$h^{-1}$Mpc, where the
minimum distance comes from requiring a reliable redshift distance
estimate (relatively unaffected by peculiar velocities) and the maximum
distance is due to the magnitude limits of the survey. We assume a
$q_{0} = \frac{1}{2}$, $\Lambda = 0$ cosmology and so comoving
distances are given by
\begin{equation}
r(z) = \left( \frac{2c}{H_{0}} \right) \left[ 1 - \frac{1}{\sqrt{1+z}}
\right] ,
\end{equation}
where $H_{0}=100h$ kms$^{-1}$Mpc$^{-1}$ is the Hubble constant and $c$
is the velocity of light in kms$^{-1}$. Volumes are calculated from the
comoving distances and use
\begin{equation}
V = \frac{d\Omega}{3} r^{3}(z) ,
\end{equation}
where $d\Omega$ is the solid steradian angle of the survey. Absolute
($M$) and apparent ($m$) magnitudes are related by
\begin{equation}
5\lg d_{L}(z) = m - M - 25 - k_{corr}(z) ,
\end{equation}
where 
\begin{equation}
d_{L}(z) = (1+z) \; r(z) ,
\end{equation}
is the luminosity distance in $h^{-1}$Mpc and $k_{corr}$ is the
k-correction. A simple k-correction is used for all galaxies
independent of morphological type
\begin{equation}
k_{corr} = k_{1}z + k_{2}z^{2} ,
\end{equation}
where $k_{1} = +3.15$ and $k_{2} = -0.29$ \cite{kcorr}. Given the
redshift range of interest this k-correction is more than adequate for
this analysis.

The survey's magnitude limits were chosen by the authors to maximize
depth and minimize observational incompleteness. In this case each of
the 60 UKST fields has a different magnitude limit and our best sample
contains 2055 redshifts with $\left< m_{lim} \right> = 16.86 \pm 0.25$
mags and an average completeness of 75 per cent. Selection effects like
this have been shown to cause no systematic biases in our methods of
analysis \cite{myphd}. One can check the completeness of this sample by
using Schmidt's (1968) $\left< V/V_{max} \right>$ test. For a uniform
distribution of galaxies in a complete magnitude limited sample $\left<
V/V_{max} \right> = 0.5$, with a rms dispersion of $1/\sqrt{12N}$ (in
the absence of clustering). If $\left< V/V_{max} \right>$ is
significantly lower or higher than 0.5 then we are missing objects at
high or low redshift, respectively. For the above best sample we find
that $\left< V/V_{max} \right> = 0.50 \pm 0.01$ and so we are not
systematically missing galaxies at any redshift.

\subsection{The Luminosity Function Shape} \label{parasec}

\begin{figure*}
\centering
\centerline{\epsfxsize=17.0truecm \figinsert{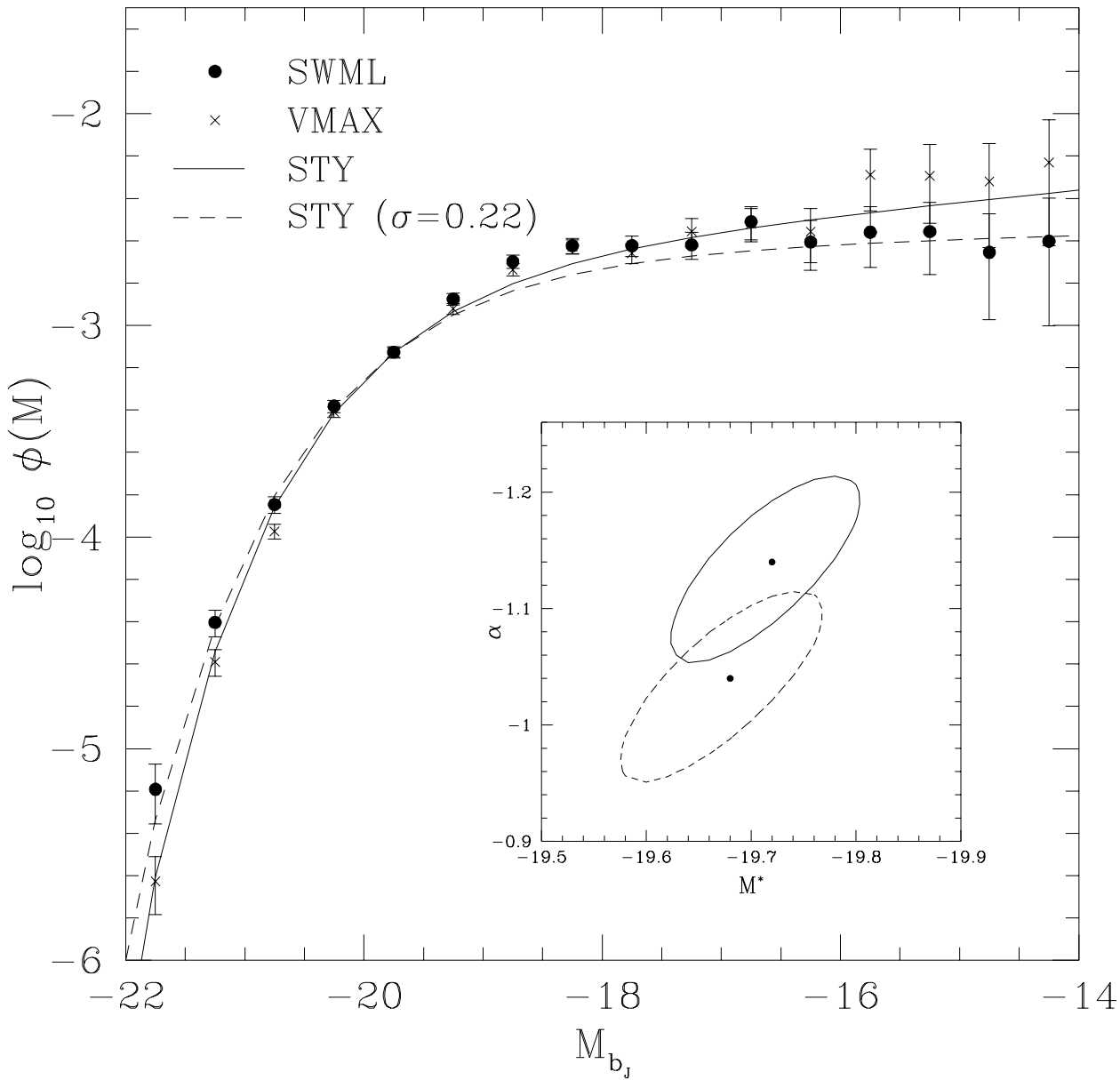}{0.0pt}}
\caption{The galaxy luminosity function maximum likelihood solutions
for a stepwise function (dots), a pure Schechter function (solid curve)
and a convolved Schechter function (dashed curve). Also shown for
comparison is the basic VMAX luminosity function estimate (crosses).
These four solutions are scaled to agree at $M_{b_{J}}=-19.75$ but the
overall normalisation is arbitrary at this stage. The inset shows the
STY maximum likelihood results and joint 68 per cent error ellipsoids
on both parameters in the ($M^{*} , \alpha$) plane.} \label{schfig1}
\end{figure*}

\begin{table}
\begin{center}
\caption{STY maximum likelihood results and normalisations assuming the
pure Schechter function of equation~\ref{sch}, $\phi$, and the
convolved Schechter function of equation~\ref{consch}, $\phi_{o}$.}
\label{scstab}
\begin{tabular}{ccc}
& {\Large $\phi$} & {\Large $\phi_{o}$} \\
& & \\
$\alpha$ & $-1.14 \pm 0.08$ & $-1.04 \pm 0.08$ \\
$M^{*}_{b_{J}}$ ($h = 1$) & $-19.72 \pm 0.09$ & $-19.68 \pm 0.10$ \\
Prob. & $0.16$ & $0.22$ \\
$\bar{n}$ ($h^{3}$Mpc$^{-3}$) & $(6.4 \pm 1.1) \times 10^{-2}$ & $(4.4 \pm
0.8) \times 10^{-2}$ \\
$\phi^{*}$ ($h^{3}$Mpc$^{-3}$) & $(1.2 \pm 0.2) \times 10^{-2}$ & $(1.7
\pm 0.3) \times 10^{-2}$ \\
\end{tabular}
\end{center}
\end{table}

Using the magnitude limits from the previous section the STY parametric
solution has been calculated from the Durham/UKST survey for a pure
Schechter function (equation~\ref{sch}) and a convolved Schechter
function (equation~\ref{consch}). Following Metcalfe et al. (1995b) we
use $\sigma = 0.22$ in equation~\ref{consch} for the rms scatter in the
observed magnitudes. Small variations in this parameter do not
significantly affect the final results. The maximum likelihood results
for $\alpha$ and $M^{*}$ are shown in Table~\ref{scstab} and assume
$h=1$. The SWML non-parametric solution has similarly been calculated
from the Durham/UKST survey, taking $\sim$20 iterations for 5 $s.f.$
convergence. These maximum likelihood solutions are all shown in
Fig.~\ref{schfig1}, along with the basic VMAX estimate of the
luminosity function, where we have scaled the $\phi$'s to agree in the
bin containing the most galaxies ($M_{b_{J}}=-19.75$). We are only
considering the shape of the luminosity function here and so the
absolute normalisation is still arbitrary. The inset of
Fig.~\ref{schfig1} shows the joint 68 per cent error ellipsoids for the
two STY solutions as calculated from equation~\ref{styerr}, while the
errors quoted in Table~\ref{scstab} are the $1 \sigma$ error on an
individual parameter. The VMAX and SWML error bars on
Fig.~\ref{schfig1} were calculated from equations~\ref{vmaxerr}
and~\ref{swmlerr}, respectively.

The observational incompleteness described in Section~\ref{redsursec}
is not explicitly accounted for in the luminosity function shape
analysis presented here. To get around this problem one can weight the
$p_{i}$'s of equation~\ref{prob} by the appropriate completeness rate
depending on the UKST field and apparent magnitude interval in
question. In practice, this correction made very little or no
difference to the either the STY or the SWML maximum likelihood
solutions.

A comparison of the two STY solutions in Fig.~\ref{schfig1} shows that
the main effect of the magnitude errors on a luminosity function of
this shape is to pull $\phi(M)$ down at faint magnitudes while pushing
it up slightly at bright magnitudes, essentially flattening it
overall. The VMAX and SWML estimates both agree well except at the
very bright and very faint ends. Such a faint end excess is expected if
there local overdensity, see Fig.~\ref{denfig1}, and rejecting all
galaxies within $r < 20h^{-1}$Mpc brings the VMAX estimate into line
with the SWML one. The STY and SWML estimates both have relatively
flat faint end slopes with no convincing evidence for a faint end
upturn down to $M_{b_{J}} \sim -15$. The results of the likelihood
ratio test described in Section~\ref{lumfnmethsec} are also given in
Table~\ref{scstab}. It is seen that neither of these Schechter
functions provide good formal fits to the SWML function, with the
convolved Schechter function being marginally better. However, both
agree well with the general SWML shape on a qualitative level.

\subsection{The Radial Density Shape} \label{raddenshape}

\begin{figure}
\centering
\centerline{\epsfxsize=8.5truecm \figinsert{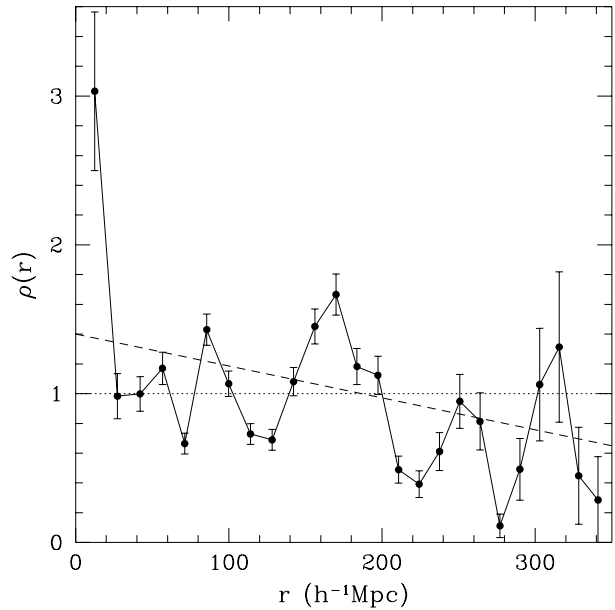}{0.0pt}}
\caption{The maximum likelihood estimate for the radial density
function. The solution has been normalised to unity in the observed
distance range as indicated by the dotted line. Large fluctuations of
order $\sim$50 per cent are present on $\sim$50$h^{-1}$Mpc scales.
Also, the dashed line shows a simple straight line minimum $\chi^{2}$
fit to the data points. This confirms the visual impression that the
radial density is falling with distance.} \label{denfig1}
\end{figure}

The maximum likelihood estimate of the radial density function has been
calculated using the method outlined in Section~\ref{raddenmethsec} and
is shown in Fig.~\ref{denfig1}. The error bars on $\rho(r)$ were also
calculated using the techniques described in
Section~\ref{raddenmethsec}. The non-parametric stepwise solution
converged to \mbox{5 $s.f.$} after $\sim$20 iterations. This solution
has been normalised to unity over the plotted distance range.

The observational incompleteness previously described is not explicitly
corrected for in this method. One can again weight the $p_{i}$'s
implicit in equation~\ref{raddeneqn} by the appropriate completeness
rate to account for this effect. In practice, this correction makes
only a small difference to the estimated $\rho(r)$.

Fig.~\ref{denfig1} shows that fluctuations in the observed radial
density function are of order $\sim$40-60 per cent and occur on
$\sim$50$h^{-1}$Mpc scales. As we will see the radial distances of the
`peak and trough' fluctuations agree well with the observed
number-distance histogram. The observed large local overdensity for
$r<20h^{-1}$Mpc probably causes the slight faint end excess seen in the
VMAX estimate of the luminosity function. Also, it is interesting to
see that a straight line minimum $\chi^{2}$ fit to the radial density
function (dashed line) is significantly different from the expected
flat slope (dotted line). (This difference remains significant even if
we remove the first point from the fit.) Specifically, the $<
200h^{-1}$Mpc region is $\sim$20 per cent overdense and the $>
200h^{-1}$Mpc region $\sim$20 per cent underdense, the error being
$\sim$4 per cent on each measurement. This is discussed further in
Section~\ref{disssec}.

\begin{figure}
\centering
\centerline{\epsfxsize=8.5truecm \figinsert{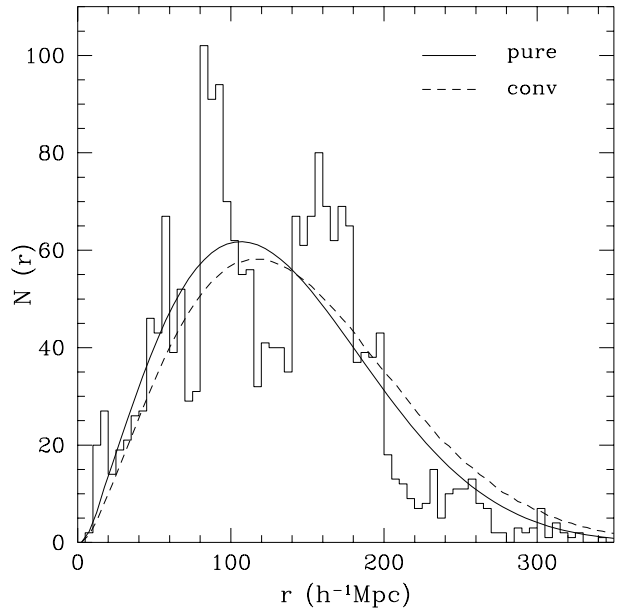}{0.0pt}}
\caption{The observed galaxy number-distance histogram, $N(r)$, from
the Durham/UKST survey using the magnitude limits described in
Section~\ref{ressec}. The smooth curves show the expected distribution
for a homogeneous sample with this survey's magnitude limits and
completeness rates using the maximum likelihood pure Schechter function
(solid curve) and convolved Schechter function (dashed curve) of
Section~\ref{parasec}.}
\label{nzfig}
\end{figure}

Fig.~\ref{nzfig} compares the observed galaxy number-distance histogram
with two homogeneous models. The magnitude limits and completeness
rates from the previous section are used for models and data alike. The
two models are the STY maximum likelihood solutions from
Table~\ref{scstab} for a pure (unconvolved) Schechter function and a
convolved Schechter function. They have been normalised to match the
total galaxy number in the sample. These homogeneous models do not
provide a detailed fit to the data because of the effects of strong
galaxy clustering on scales $>20h^{-1}$Mpc. It can be also seen that
both models overpredict the numbers of galaxies at $r > 200h^{-1}$Mpc,
with the convolved Schechter function model being slightly worse in
this respect.

\begin{figure}
\centering
\centerline{\epsfxsize=8.5truecm \figinsert{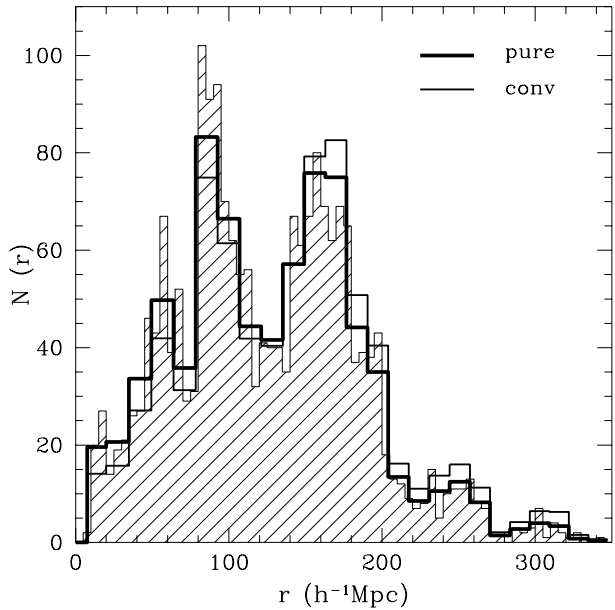}{0.0pt}}
\caption{The observed galaxy number-distance histogram, $N(r)$, from
the Durham/UKST survey (shaded histogram, see Fig~\ref{nzfig}). The two
bold histograms are the product of the Durham/UKST radial density
function (see Fig.~\ref{denfig1}) and the two number-distance model
curves (see Fig~\ref{nzfig}).} \label{nzfig2}
\end{figure}

Comparison of the observed $N(r)$ from Fig.~\ref{nzfig} with the radial
density function from Fig.~\ref{denfig1} shows that the `peak and
trough' fluctuations in $\rho(r)$ agree well with the `spikes' in
$N(r)$. As a consistency test, in Fig.~\ref{nzfig2} we plot the product
of the two homogeneous $N(r)$ models with the observed $\rho(r)$ and
compare with the observed $N(r)$. The agreement is impressive, implying
that the initial assumption of the luminosity function having a
universal form (allowing a separation of variables, see
Section~\ref{lumfnmethsec}) is justified.

\subsection{The Normalisation}

The results of the iterative estimator of equations~\ref{nbar3}
and~\ref{weight} for $\bar{n}$ are given in Table~\ref{scstab}, with
$\sim$5 iterations needed for 5 $s.f.$ convergence. The value of
$\phi^{*}$ is estimated from numerical evaluation of the integral in
equation~\ref{nbar2}. Use of a simple \mbox{$w = 1$} estimate for
$\bar{n}$ (instead of equation~\ref{weight}) makes a $\sim$5 per cent
difference to $\bar{n}$. The formal error from equation~\ref{nbarerr}
(using the weighting in equation~\ref{weight}) gives a 7 per cent error
in $\bar{n}$. This can be compared with a 10 per cent error for
\mbox{$w = 1$.} However, the dominant error comes from the uncertainty
in the selection function (due to the uncertainty in the Schechter
parameters $M^{*}$ and $\alpha$) and is $\sim$15 per cent. This gives a
total error in the normalisation of $\sim$17 per cent. Also, for this
analysis we choose $M_{low} = -15$ and as previously mentioned in
Section~\ref{normmethsec} varying this parameter only causes a small
($\sim$ few per cent) change in $\phi^{*}$.

Obviously for estimates of $\bar{n}$ and $\phi^{*}$ the effects of
observational incompleteness must be corrected. Therefore, in this
analysis the sum in the numerator of equation~\ref{nbar2} is weighted
by the appropriate completeness rate of the UKST field and apparent
magnitude interval of the galaxy in question. 

One can comment on the $\sim$40 per cent change in $\phi^{*}$ observed
with the two different parametric forms. This is caused by a
combination of the different relative luminosity function shapes (and
the corresponding integrals over them in
equations~\ref{selfn},~\ref{nbar2} and~\ref{nbar3}) and the different
Schechter function parameters (particularly $\alpha$). This is
discussed further in Section~\ref{disssec}.

\section{Discussion} \label{disssec}

\subsection{Comparison with Luminosity Functions from Other Galaxy
Redshift Surveys} \label{complfnsec}

\begin{figure}
\centering
\centerline{\epsfxsize=8.5truecm \figinsert{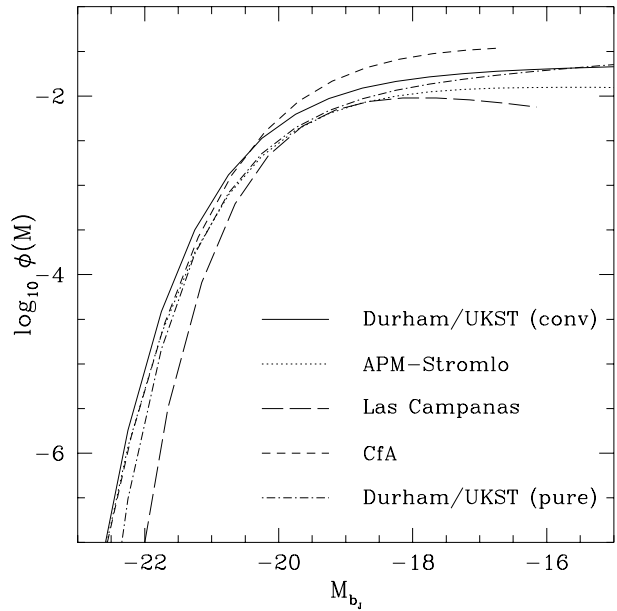}{0.0pt}}
\caption{Comparison of the normalised Durham/UKST luminosity function
with those estimated from recent galaxy redshift surveys. All the
magnitudes have been transformed to the blue $b_{J}$ system. This
entails shifting the Las Campanas Gunn-$r$ magnitudes by the average
galaxy colour, $\left< b_{J} - r \right>_{0} = +1.1$ (Tucker 1994; Lin
et al. 1996), while the CfA Zwicky magnitudes are shifted by $b_{J} -
M_{Z} = -0.45$ (Shanks et al. 1984). All of the luminosity functions
plotted here are corrected for magnitude errors by convolution of a
Gaussian error function with a Schechter function. For reference, we
also plot the pure (unconvolved) Durham/UKST Schechter function.}
\label{compfig}
\end{figure}

\begin{table*}
\begin{center}
\caption{Comparison of some recent galaxy redshift survey parameters
and their corresponding maximum likelihood (convolved) Schechter
luminosity functions. While the original passband of the survey is
quoted in the left-hand column, the $M^{*}$'s in the right-hand column
have been transformed to the $b_{J}$ passband so that a direct
comparison can be made ($\left< b_{J} - r \right>_{0} = +1.1$ and
$b_{J} - M_{Z} = -0.45$, Tucker 1994; Lin et al. 1996; Shanks et al.
1984).} \label{comptab}
\begin{tabular}{ccccccc}
Survey & Passband & $N_{gal}$ & Volume ($h^{-3}$Mpc$^{-3}$) & $\alpha$
& $M_{b_{J}}^{*}$ & $\phi^{*}$ ($h^{3}$Mpc$^{-3}$) \\
& & & & & & \\
Durham/UKST & $b_{J}$ & $\sim$2100 & $4 \times 10^{6}$ & $-1.04 \pm
0.08$ & $-19.68 \pm 0.08$ & $(1.7 \pm 0.3) \times 10^{-2}$ \\
APM-Stromlo & $b_{J}$ & $\sim$1800 & $1 \times 10^{7}$ & $-0.97 \pm
0.15$ & $-19.50 \pm 0.13$ & $(1.4 \pm 0.2) \times 10^{-2}$ \\
Las Campanas & Gunn-$r$ & $\sim$19000 & $1 \times 10^{7}$ & $-0.70 \pm
0.05$ & $-19.19 \pm 0.02$ & $(1.9 \pm 0.1) \times 10^{-2}$ \\
CfA & Zwicky & $\sim$9000 & $1 \times 10^{6}$ & $-1.0 \pm 0.2$ &
$-19.25 \pm 0.3$ & $(4.0 \pm 1.0) \times 10^{-2}$ \\
\end{tabular}
\end{center}
\end{table*}

Table~\ref{comptab} shows a comparison between the parameters of some
recently completed galaxy redshift surveys (Loveday et al. 1992; Marzke
et al. 1994; Lin et al. 1996) and the Durham/UKST survey. Note that the
comparisons given here are for {\it convolved} Schechter functions
because they are the preferred fits as quoted in the literature.
However, as was seen in Section~\ref{parasec}, while the convolved
Schechter function did provide a slightly better fit to the SWML
luminosity function, the significance was marginal given the extra
complexity added to the analysis. As was mentioned in
Section~\ref{raddenshape} the pure (unconvolved) Schechter function
provides a more realistic fit to the observed number-distance
histogram. These normalised luminosity functions are plotted in
Fig.~\ref{compfig} where magnitudes have been transformed to the blue
$b_{J}$ system, albeit with only a naive magnitude offset. The Las
Campanas galaxies are offset by the average galaxy colour, namely
$\left< b_{J} - r \right>_{0} = +1.1$ (Tucker 1994; Lin et al. 1996).
The CfA galaxies are offset by $b_{J} - M_{Z} = -0.45$, which comes
from the comparison of galaxy number-magnitude count data in Shanks et
al. (1984).

Looking at Table~\ref{comptab} shows that the Las Campanas and CfA
surveys contain considerably more galaxies than the Durham/UKST or
APM-Stromlo surveys. However, in spite of this, all of these surveys
sample similar sized volumes (within an order of magnitude) due to the
different observing strategies used. The accuracy of the quoted
$\phi^{*}$'s reflects both the number of galaxies and, more
importantly, the total volume surveyed.

Comparing the results in Table~\ref{comptab} and Fig.~\ref{compfig}
shows that the general features of these luminosity functions agree
well down to $M_{b_{J}} \sim -17$. Namely, that the characteristic
absolute magnitude at the `knee' is $M_{b_{J}}^{*} \simeq -19.4$, the
faint end slope is flat, $\alpha \simeq -1.0$, and the overall
normalisations agree well (apart from the CfA one which may be biased
high by local inhomogeneities). However, fainter than $M_{b_{J}} \sim
-17$ one does find some discrepancies in the values of the faint end
slopes. The Durham/UKST and APM-Stromlo surveys agree well and retain a
reasonably flat slope. Marzke et al. (1994) claim that the CfA survey
has a significant excess of galaxies in this region, $\alpha \simeq
-1.3$. Lin et al. (1996) measure a declining faint end slope, $\alpha
\simeq -0.7$, from the Las Campanas survey.
 
We can comment on the results from these other surveys. Firstly, the
CfA excess comes mainly from a very local region ($r < 25h^{-1}$Mpc),
which is probably not a representative sample of the Universe.
Therefore, while an excess might exist, its significance could be in
doubt. Secondly, not only is the Las Campanas survey selected in a
different passband (which would cause different relative fractions of
galaxy types to be observed), but also this survey is biased against
observing low surface brightness galaxies (because of observational
selection effects). Therefore, if there is any correlation between low
surface brightness and intrinsically faint absolute magnitudes then one
can explain their deficit. We also comment on the preliminary
luminosity function results from the ESO Slice Project (ESP; Vettolani
et al. 1996). They measure a normalisation approximately twice that of
the above surveys and a rising faint end slope, $\alpha \simeq -1.2$,
for galaxies fainter than $M_{b_{J}} \sim -17$. At present the
significance of their faint end excess is unknown and so we cannot
comment on that result. However, given that this survey is complete to
a fainter magnitude limit than those discussed here, $b_{J} \simeq
19.5$, one might expect a higher normalisation because of the steep
slope of the observed galaxy number-magnitude counts in this range
(e.g. Metcalfe et al. 1995a).

We also note that the form of the Durham/UKST and APM-Stromlo
luminosity functions are in good agreement with the combined
colour-dependent galaxy luminosity function estimates of Shanks (1990)
and Metcalfe et al. (1997). Thus, the increasing  slope of the
luminosity function of blue, late-type galaxies cancels with the
decreasing luminosity function slope of the redder galaxies, leaving an
overall luminosity function which has a flat faint end, which is
observed here.

Finally, it is possible to be  too pessimistic about the statistical
robustness of luminosity function estimates from local magnitude limited
surveys (e.g. Driver \& Phillipps 1996). We emphasise again that the
Durham/UKST luminosity function presented here agrees very well with that
of the APM-Stromlo survey, while the observed differences in the other
surveys could be caused by slight systematic problems.

\subsection{The Radial Density Distribution} \label{nzhistsec}

\begin{figure}
\centering
\centerline{\epsfxsize=8.5truecm \figinsert{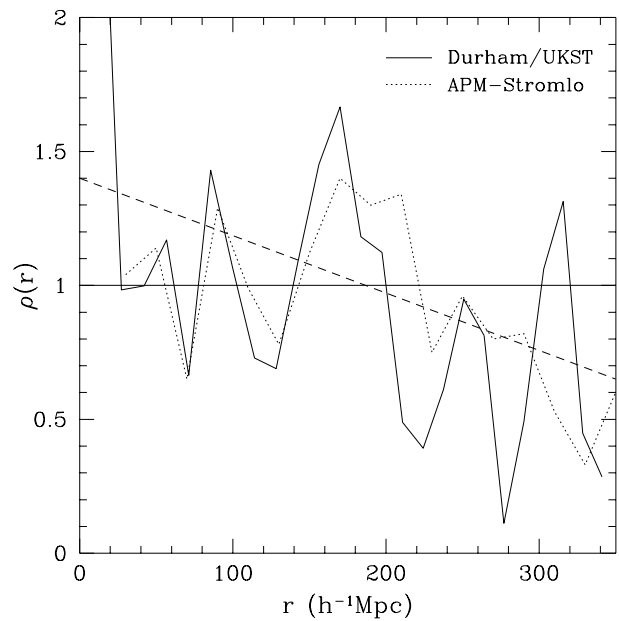}{0.0pt}}
\caption{Comparison of the radial density functions, $\rho(r)$,
estimated from the Durham/UKST and the APM-Stromlo surveys. These two
surveys map a similar overlapping Southern region of space, with the
Durham/UKST survey sampling a $\sim$1500 sq. deg. area and the
APM-Stromlo survey a $\sim$4300 sq. deg. area. These two solutions have
been normalised to unity within each survey. The dashed line shows the
minimum $\chi^{2}$ straight line fit to the Durham/UKST data from
Fig.~\ref{denfig1}.} \label{denfig2}
\end{figure}

Fig.~\ref{denfig2} gives a comparison of the radial density functions,
$\rho(r)$, for the Durham/UKST and APM-Stromlo surveys. These surveys
overlap each other in a Southern region of the sky but map the space
slightly differently. The Durham/UKST survey maps a `wedge'
($\sim$$20^{\circ} \times 75^{\circ}$) with 1 in 3 sparse sampling,
while the APM-Stromlo survey maps a thicker `wedge' ($\sim$$55^{\circ}
\times 80^{\circ}$) with 1 in 20 sparse sampling. These two solutions
have been normalised to unity within each survey. It can be seen that
there is good agreement between the two observed $\rho(r)$'s despite the
fact that they do sample slightly different regions of space. In
particular, the `peak and trough' fluctuations are observed in roughly the
same places for both surveys. Also, the falling radial density seen in the
Durham/UKST survey again appears in the APM-Stromlo survey. 

It has previously been suggested that a very large local void, $>
100h^{-1}$Mpc in extent, could explain the low normalisation and steep
slope seen in the bright galaxy number-magnitude counts ($b_{J} <$ 17),
without the need for evolution (e.g. Shanks 1990). The observed trend
in $\rho(r)$ lends support to the idea that large-scale structure {\it
can} affect the counts over a wide magnitude range.  However, it should
be noted that the decreasing trend in $\rho(r)$ seen in
Fig.~\ref{denfig2} extends over the whole range of these surveys, out
to $z \simeq 0.1$. Since the counts beyond $b_{J} = 17$ rise to have a
$2\times$ higher $\phi^{*}$ than at brighter magnitudes this means that
$\rho(r)$ will have to rise sharply just beyond the $z \simeq 0.1$
range of the present survey if a large scale structure explanation of
the steep galaxy counts is to be possible.

The alternative possibility that the steep counts slope is caused by
galaxy luminosity evolution is constrained by the fact that, at $b_{J} <
21$, the galaxy $M^{*}$ seems unevolved, which confines any evolutionary
explanation to affecting only the intrinsically faint galaxies
(Broadhurst, Ellis \& Shanks 1988). The observed {\it decrease} of
$\rho(r)$ with distance means that any such evolutionary increase in
the numbers of faint galaxies  has to be even larger in the range $0 <
z < 0.1$ than previously expected to explain the increase in galaxy
counts at $17 < b_{J} < 19$.

We have also considered the possibility that the decreasing radial
density $\rho(r)$ could be caused by the effect of cosmology on the
volume element. As noted by Weinberg (1972), the effect of $q_{0}$ on
the volume element with redshift is first order, $dV \propto
(1-{3\over2}q_{0}z)dz$; thus even at $z = 0.1$ there is a 15\% change
in volume for a unit change in $q_{0}$.  However, the $\rho(r)$
displayed in Fig.~\ref{denfig2} was calculated using $q_{0}=0.5$.
Assuming a lower value of $q_{0}$ such as $q_{0}=0.05$ or $q_{0}=-0.5$
would therefore increase the volume element at high redshift and
decrease the density at high redshift still further, acting in the
opposite manner to that required to reduce the redshift dependence of
$\rho(r)$.  Thus only a significantly higher value of $q_{0}$ than
$q_{0}=0.5$ would produce a more homogeneous result and we regard this
as less likely than the alternative large-scale structure or
evolutionary explanations.

\subsection{Comparison with the Cluster Galaxy Luminosity Function}

All of the galaxy luminosity functions mentioned so far have been
estimated from the field. We can qualitatively compare our results (and
those from other redshift surveys) with estimates of the galaxy
luminosity function from clusters.

A faint end excess in the cluster galaxy luminosity function ($\alpha
\simeq -1.4$) has been known to exist in the Coma and Virgo clusters
for quite some time (e.g. Abell 1977; Metcalfe 1983; Binggeli, Sandage
\& Tammann 1985). Recent work has confirmed these initial results and
extended the observations to other clusters (e.g. Biviano et al. 1995;
Kashikawa et al. 1995). Driver et al. (1994) and Wilson et al. (1996)
have also studied clusters at higher redshifts ($z \sim 0.2$) and
observe a similar steeply rising upturn at the faint end. The overall
conclusion is that a single Schechter function cannot provide an
adequate fit to such luminosity functions. In all of this recent work
the cluster galaxy luminosity function appears better fit by a
2-component Schechter-like function with a flat bright slope ($\alpha
\sim -1$) and a steep upturn ($\alpha \sim -2$) for galaxies fainter
than $M^{*} + 2$.

\begin{figure}
\centering
\centerline{\epsfxsize=8.5truecm \figinsert{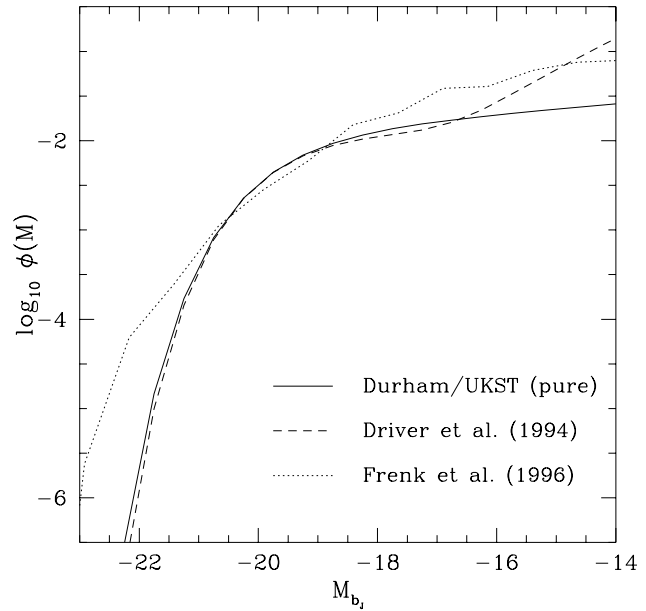}{0.0pt}}
\caption{Comparison of the normalised Durham/UKST pure Schechter
luminosity function with the cluster galaxy luminosity function model
of Driver et al. (1994) and the galaxy formation model of Frenk et al.
(1996). The model of Driver et al. (1994) has been made to agree with
the Durham/UKST results at $M^{*}$ and $\phi^{*}$. For the Frenk et al.
(1996) results the normalisations are fixed by the model.}
\label{compfig2}
\end{figure}

In Fig.~\ref{compfig2} we compare the STY results for the pure
Schechter function of Section~\ref{parasec} with the model proposed by
Driver et al. (1994). This model approximates the general cluster
luminosity function discussed above and has been translated in the
$M_{b_{J}}$ and $\phi$ direction to match the Durham/UKST results in
$M^{*}$ and $\phi^{*}$. As might be expected (given the above
discussion) this form does not give a particularly good representation
of the Durham/UKST results at the faint end of the luminosity function
(similarly for the APM-Stromlo and Las Campanas luminosity functions).
On the other hand, it is tempting to say that the results from the CfA
and ESP surveys match this form of luminosity function. Overall, the
errors in the faint ends of both the cluster and field luminosity
functions are probably large enough such that no significant
inconsistencies currently exist between the two functions.

\subsection{Comparison with Current Models of Galaxy Formation}

One can also qualitatively compare our results with those from current
models of galaxy formation and evolution. These models approach the
galaxy formation problem by combining the results of
N-body/hydrodynamic simulations with semianalytic modelling techniques
(e.g. Cole et al. 1994).

In order to successfully model galaxy formation one needs to include
many distinct physical processes, as well as an underlying cosmological
structure formation model. These processes include the evolution of
dark matter halos, the dynamics of gas cooling, star formation and
feedback, the evolution of the stellar populations that form and how
galaxies merge and interact. One approximates these processes by a set
of simple rules which then form the basis of a semianalytic model of
galaxy formation (e.g. Kauffmann et al. 1993; Lacey et al. 1993; Cole
et al. 1994). The models are then used to predict and compare with the
observable properties of the galaxy distribution, such as the faint
galaxy number-magnitude counts, galaxy colours and the galaxy
luminosity function.

Fig.~\ref{compfig2} also compares the pure Schechter function STY
results from Section~\ref{parasec} with the semianalytic galaxy
formation model of Frenk et al. (1996). The details of this fiducial
model are described in Cole et al. (1994) with recent updates from
Frenk et al. (1996) and Baugh et al. (1996). One can see that, given
the complex modelling involved, this is a reasonable approximation to
the luminosity function around $M^{*}$.  Note that the excess of very
bright galaxies has been previously documented (e.g. Frenk et al.
1996). Also, the model predicts a steep faint end slope to the
luminosity function. Although this is more in line with the faint end
excess seen in the CfA and ESP surveys, we have argued that these are
less reliable estimates of the {\it local} galaxy luminosity function,
which we believe is better represented by the results from the
Durham/UKST, APM-Stromlo and Las Campanas surveys.

\section{Conclusions} \label{concsec}

We have estimated the galaxy luminosity function from the Durham/UKST
Galaxy Redshift Survey. We use standard maximum likelihood techniques
that are unbiased by density inhomogeneities and an optimal weighting
function for estimating the normalisation. For a standard Schechter
function we find $M_{b_{J}}^{*} = -19.72 \pm 0.09$, $\alpha = -1.14 \pm
0.08$ and $\phi^{*} = (1.2 \pm 0.2) \times 10^{-2} h^{3}$Mpc$^{-3}$.

Correcting for the observed scatter in our measured magnitudes causes a
flatter faint end slope to be measured ($\alpha = -1.04 \pm 0.08$),
although  this does not significantly improve the quality of the fit. A
combination of the change in the shape of the functional form and the
flatter $\alpha$ then causes a higher normalisation to be estimated,
$\phi^{*} = (1.7 \pm 0.3) \times 10^{-2} h^{3}$Mpc$^{-3}$.

Comparison with other recent estimates of the galaxy luminosity
function from redshift surveys gives good agreement for absolute
magnitudes brighter than $M_{b_{J}} \sim -17$. Fainter than this
absolute magnitude there are some discrepancies in the measured value
of $\alpha$, ranging from -0.7 to -1.3. The Durham/UKST survey lies
approximately in the middle of this range and is in very good agreement
with the results from the APM-Stromlo survey.

The radial density function has been estimated from the Durham/UKST
survey using a maximum likelihood technique. This shows evidence for
a falling galaxy density with radial distance, suggesting that
large-scale structure may affect the form of the galaxy number
magnitude counts in the range $14<b_J<19$.

We have shown that the result of forming the product of the radial
density function with homogeneous models which assume the above
luminosity functions gives good agreement with the observed
number-distance histogram. This result is consistent with the initial
assumption that the galaxy luminosity function has a universal form.

Our luminosity function results for the field agree with those derived
from galaxies in clusters, except at the faint end where the cluster
luminosity function appears to be steeper.

Finally, we have compared our results with the predictions of current
models of semi-analytic galaxy formation and evolution. These also
disagree with our luminosity function  at the faint end, where the
galaxy formation models predict a significantly  steeper luminosity
function slope.

\section*{acknowledgments}

We are grateful to the staff at the UKST and AAO for their assistance
in the gathering of the observations. L. Teodoro and C.M. Baugh are
thanked for useful discussions. AR acknowledges the receipt of a PPARC
Research Studentship and PPARC are also thanked for allocating the
observing time via PATT and for the use of the STARLINK computer
facilities.

\end{document}